\documentclass[american,aps,pra,reprint, superscriptaddress]{revtex4-1}
\usepackage{color}
\usepackage{times}
\usepackage{subfig}
\usepackage{bm}
\usepackage{graphicx}
\usepackage{graphics}
\DeclareGraphicsExtensions{.pdf,.png,.jpg}
\usepackage{amsbsy}
\usepackage{amsmath}
\usepackage{amsfonts}
\usepackage{amsthm}
\usepackage{float}
\usepackage{amssymb}
\usepackage{textcomp}
\usepackage{mathtools}
\usepackage[subnum]{cases}
\usepackage{enumitem}

\DeclareMathOperator{\Tr}{Tr}
\definecolor{purple(html/css)}{rgb}{0.5, 0.0, 0.5}
\newcommand{\ket}[1]{| #1 \rangle}

\newcommand{\braket}[2]{\langle #1 | #2 \rangle}
\newcommand{\ketbra}[2]{| #1 \rangle \langle #2 |}

\begin{document}
	\title{Role of fine-grained uncertainty in determining the limit of preparation contextuality
	}
	\author{Gautam Sharma}
	\email{gautam.oct@gmail.com}
	\affiliation{Quantum Information and Computation Group, Harish-Chandra Research Institute,\\ Homi Bhabha National Institute, Allahabad, 211019, India}
	
	\author{Sk Sazim}
	\email{sk.sazimsq49@gmail.com}
	\affiliation{Quantum Information and Computation Group, Harish-Chandra Research Institute,\\ Homi Bhabha National Institute, Allahabad, 211019, India}
	
		\author{Shiladitya Mal}
	\email{shiladitya@hri.res.in}
	\affiliation{Quantum Information and Computation Group, Harish-Chandra Research Institute,\\ Homi Bhabha National Institute, Allahabad, 211019, India}

	\begin{abstract}
             The optimal success probability of a communication game sets fundamental limitations on an operational theory. Quantum advantage of parity oblivious random access code (PORAC), a communication game, over classical resources reveals the preparation contextuality of quantum theory [Phys. Rev. Lett. {\bf{102}}, 010401 (2009)]. Optimal quantum advantage in the N-dit PORAC game for finite dimensions is an open problem. Here, we show that the degree of uncertainty allowed in an operational theory determines the amount of preparation contextuality. We connect the upper bound of fine-grained uncertainty relation to the success probability of PORAC game played with the quantum resource. Subsequently, we find the optimal success probability for the 2-dit PORAC game using MUBs for the decoding strategy. Finally, we also derive an upper bound on quantum advantage for the N-dit PORAC game.
	\end{abstract}
	
	\maketitle
	\section{Introduction}
Quantum physics has several fundamental no go theorems, which reveals how radically it deviates from classical physics. Bell theorem states that quantum theory cannot be reproduced by a local realist model \cite{bell,brunner}. On the other hand, the  Bell-Kochen-Specker theorem asserts that quantum theory is contextual \cite{Kochen1967-KOCTPO-3,RevModPhys.38.447}. It  means that the observables cannot be assigned definite values, independent of the setting in which they are measured, i.e., the context. Later, the notion of contextuality was generalized so that it can be associated with any operational theory \cite{PhysRevA.71.052108}. 

These no go theorems arise out of quantum correlations\cite{brunner}. In the context of spatial correlation, it is known that nonlocality of a theory is not enough so that it allows signaling \cite{PRbox}. Quantum correlation between space-like separated measurements is restricted by the Cirelson type bound \cite{Cirel'son1980}. Subsequently, it was asked whether there are some physical principles which limit the amount of nonlocality. There are approaches from information theory \cite{VerSteeg:2009:RUR:2011804.2011810,2009Natur.461.1101P}, communication complexity \cite{PhysRevLett.96.250401, 2005quant.ph..1159V}, local quantum mechanics \cite{PhysRevLett.104.140401} to address this question. In Ref.\cite{Oppenheim}, the authors took a very different approach i.e., they relate the limit of nonlocality with two inherent properties of any physical theory called, uncertainty \cite{Heisenberg1927} and steerability \cite{PhysRevA.40.913,PhysRevLett.98.140402,2019arXiv190306663U}.

Initially, uncertainty relations were stated in terms of product of standard deviations lower bounded by some quantity related with commutators of the observables measured \cite{Heisenberg1927,Kennard1927}. Later, entropic uncertainty relations were introduced which are state-independent \cite{PhysRevLett.60.1103,RevModPhys.89.015002}. However, entropic measures depict uncertainty in a coarse way as it does not capture uncertainty in the realization of different outcomes distributions for multiple measurements. To circumvent this, fine-grained uncertainty relation(FUR) was introduced, which is a set of inequalities, one for each possible combination of outcomes \cite{Oppenheim}. Later this inequality was generalized for higher dimensional systems for mutually unbiased bases (MUB) \cite{Rastegin2015}. 
	
Fundamental limiting features of a theory has been often studied through the ability of some communication games to process information \cite{2009Natur.461.1101P, PhysRevA.92.052312}. Random access code (RAC) is a two-player communication game \cite{Ambainis:1999:DQC:301250.301347,2008arXiv0810.2937A}, a party, say, Alice holding a data set in the form of n-bit string, encodes it in a state and sends to another party Bob whose task is to guess any one of the bit randomly chosen from the string(see Fig.\ref{fig1}). The generalization of bits to higher dimensions is dits. A bit refers to a two-level system whereas dit is a $d$ level system. Therefore, instead of a n-bit string, Alice can also encode an n-dit string in a state and send it to Bob, who then tries to guess a dit from the string \cite{PhysRevLett.114.170502}.  
	
An interesting connection between an RAC game and contextuality was made by Spekkens et al.\cite{spekkenpom}, invoking the parity obliviousness constraint. The constraint of parity obliviousness in an RAC game demands that encoding is such that the receiver can not know the parity of the incoming signal $x$ from the sender. One of the ways of defining the parity of message $x$ is the sum (mod d) of the bit values contained in the message.  If parity obliviousness constraint is imposed on the RAC game, which we discuss in detail in the main text, then optimal success probability of winning with classical resources coincides with that when  resources are taken from noncontextual theory. Therefore, the quantum advantage of parity oblivious random access code (PORAC) game implies the preparation contextuality of quantum theory. It was also shown that preparation contextuality leads to nonlocality \cite{PhysRevLett.110.120401,manikfoundations,PhysRevA.98.032110,tavakoli2019measurement}. Following this notion connection has been made with PORAC game and other nonlocal games \cite{Ambainis2019, Chailloux_2016,PhysRevA.98.032110} and optimal quantum bound follows from the Cirelson like bound associated with nonlocality \cite{Cirel'son1980}.	
To reveal preparation contextuality, PORAC game was studied for higher-dimensional single systems and experimental realization was demonstrated as well  \cite{PhysRevLett.119.220402,spekkenpom,anwer2019noiserobust}. Optimal quantum advantage of PORAC game was derived when n-bit classical information is encoded in higher dimensional systems \cite{Chailloux_2016}. Up to a few dimension, maximal quantum violation of preparation noncontextuality inequality was also derived numerically in \cite{PhysRevLett.119.220402}. In general finding optimal quantum bound for high-level PORAC game or maximal quantum violation of non-contextuality inequality is an open question.

Here we show that the degree of uncertainty, which is a property of a theory, determines how much a theory would be preparation contextual. Specifically, we derive tight FUR for any pair of measurements in any finite dimension and show the upper bound of the FUR is closely related with the quantum advantage of PORAC game in terms of enhanced success probability over classical strategy.  We, then prove that the optimum quantum bound is reached when Alice encodes 2-dit string in single qudit state which are the maximal certain states with respect to a pair of MUBs, while Bob's decoding strategy is to perform those MUBs. We also derive the quantum upper bound of FUR for n arbitrary observables and that the FUR upper bound also gives the upper bound of the success probability in a n-dit PORAC game. Although, our upper bound for n-dit PORAC game might not be exactly reached by quantum strategy induced by the FUR. Finally, we compare some results regarding maximal quantum bound obtained previously, with our result for the sake of completeness.

The plan of the paper is as following.  In Sec.II, we describe preliminary ideas of PORAC game and parity obliviousness. Then, in Sec.III, we briefly discuss Fine-grained uncertainty relations and derive upper bounds of various set of sharp measurements. In Sec.IV, we present our main result, where we connect the FUR upper bound with the successs probability of PORAC game. We also compare our results with the known bounds. Finally, we conclude in Sec.V.

\section{Preparation Noncontextuality from Parity Oblivious Random Access Codes}
Preparation non-contextuality associated with an operational theory was first introduced in \cite{PhysRevA.71.052108}. An operational theory provides the probabilities $p(k|P,M)$ of getting an outcome $k$ given the preparation procedure $P$, and the measurement $M$. Quantum theory is also an operational theory in which a preparation procedure $P$ is represented by $\rho_P$ and a measurement is represented by a positive operator valued measure (POVM), $\Lambda_{M,k}$. The probability of getting an outcome $k$ is $p(k|P,M)=\rm Tr(\rho_P\Lambda_{M,k})$.

An operational theory is said to be preparation non-contextual if two preparations yield the same measurement statistics for all possible measurements implies probability associated with two different preparations at the hidden variable level($\lambda$) is also same, i.e,  
\begin{align}\label{pcont}
\forall M\hspace{0.1cm} \forall k;\hspace{0.1cm} p(k|P,M)=p(k|P',M) \implies p(\lambda|P)=p(\lambda|P')
\end{align}
where $\lambda$ is a hidden variable and $P$ and $P'$ denote two preparation procedures.

\begin{figure}[h]
\centering
\includegraphics[scale=0.35]{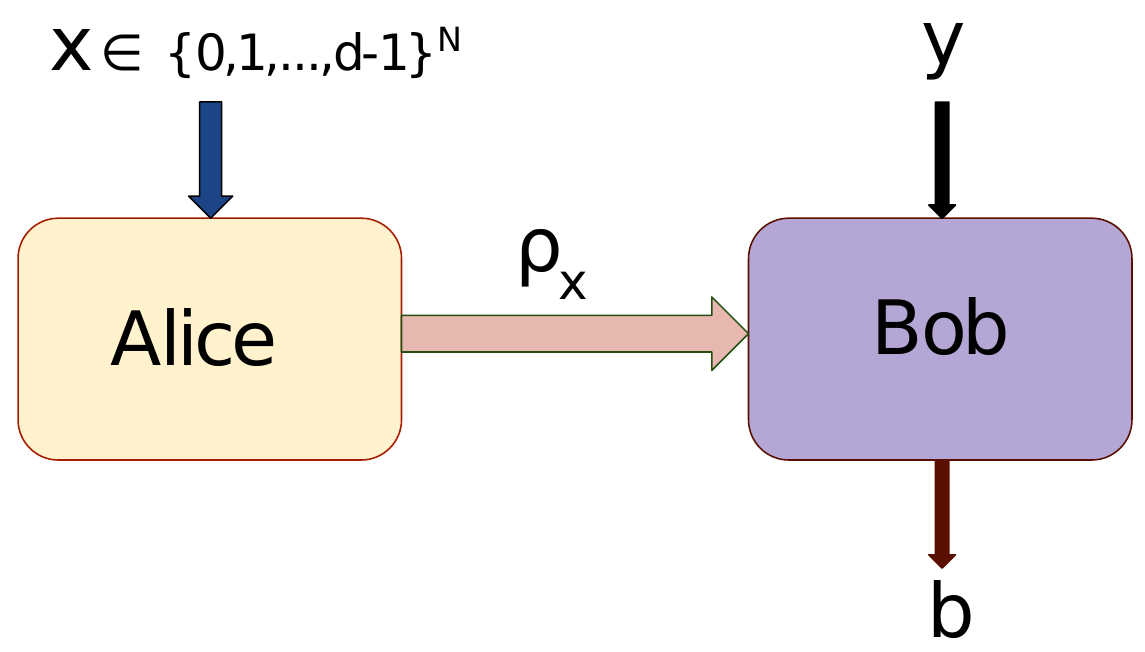}
\caption{(Color online) In this communication game, Alice encodes the classical string $x \in \{0,1,...,d-1\}^N$ in state $\rho_x$. On recieving the state $\rho_x$ Bob's performs a measurement $X_i$ chosen uniformly from a set of $N$ observables, and tries to guess the $y^{th}$ dit of  $x$ using his measurement outcome $b$.}
\label{fig1}
\end{figure}

Preparation contextuality was demonstrated using parity oblivious communication games \cite{spekkenpom,PhysRevLett.119.220402}. In the game, Alice receives an $N$-dit string  $x \in \{0,1,...,d-1\}^N$, which she encodes in a state $\rho_x$ and then sends it to Bob, chosen uniformly. Whereas, Bob's task is to guess the $y^{th}$ bit of the string $x$, using his measurement outcome $b$ obtained by a set of measurements Y, as shown in Fig.\ref{fig1}. There is a cryptographic constraint that Alice can encode her message under the parity obliviousness condition that no information about the parity of $x$ can be revealed to Bob. If $s\in Par$ where $Par\equiv\{s|s\in\{0,1,..,d-1\}^N,\zeta \leq d-2\}$, with $\zeta$ denoting the number of zeroes appearing in a particular $s$, then no information about $x\cdot s=\oplus_ix_is_i (\mod d)=l$, $\forall l\leq d-1$ should be revealed to Bob. We refer to this task as $N\rightarrow1$ $d$-Parity oblivious random access codes ($d$-PORACs). The parity obliviousness condition, for the set of measurements Y performed by Bob, can be cast down in the form of following equality
\begin{align*}
\forall s,b,l,l',y; \hspace{1mm}\frac{1}{p(l)}\sum_{x\cdot s=l}p(b|x,y)=\frac{1}{p(l')}\sum_{x\cdot s=l'}p(b|x,y),
\end{align*} 
where $p(l)=\sum_{x\cdot s=l} p(x)$. As for all $l$ parity strings $x_l$, we have $d^{N-1}$ uniform choices, $p(l)=p(l')$. Thus, the above obliviousness condition reduces to
\begin{align}\label{oblc}
\forall s,b,l,l',y; \hspace{1mm} \sum_{x\cdot s=l}p(b|x,y)=\sum_{x\cdot s=l'}p(b|x,y).
\end{align}
It should be noted that while Eq.(\ref{pcont}), is the most general form of parity obliviousness, whereas Eq.(\ref{oblc}) implies obliviousness with respect to the measurements performed by Bob. For our purpose it is sufficient to consider above condition of parity obliviousness.

Given the obliviousness constraint Bob's task is to maximize the average success probability of reporting the correct output $b=x_y$. The average probability of guessing the correct bit is given by 
\begin{align*}
p(b=x_y)=\frac{1}{d^NN}\sum_{x\in\{0,1,...d-1\}^N}\sum_{y\in\{1,...,N\}}p(b|x,y),
\end{align*}
Different operational theories provide different maximal success probability of the game. It was shown in \cite{PhysRevLett.119.220402} that an operational theory which admits a preparation non-contextual hidden-variable model, the probability of success for $N\rightarrow1$ $d$-PORAC is bounded by the following inequality,
\begin{align}\label{ncie}
\frac{1}{d^NN}\sum_{x\in\{0,1,...d-1\}^N}\sum_{y\in\{1,...,N\}}p(b|x,y)\leq\frac{N+d-1}{dN}.
\end{align}

Any operational theory which violates this inequality is contextual.


\section{Fine-Grained Uncertainty relations}
Suppose, we want to measure $N$ different observables $X_i,$ where $i\in \{1,...,N\}$, and outcomes $x_i\in\{0,...,d-1\}$. One can quantify the uncertainty associated with the measurements using entropic uncertainty relations as following
\begin{align*}
\sum_{i=1}^N H(X_i)_{\rho} \geq \beta,
\end{align*}
\noindent where $\beta$ depends on the compatibility between different observables. However, entropy is a coarse way of measuring the uncertainty and incompatibility of a set of measurements. It does not reflect the uncertainty inherent in obtaining a particular combination of outcomes $x_i$ for different measurements $X_i$. To circumvent this issue, fine-grained uncertainty relation was proposed in Ref.\cite{Oppenheim}.
The uncertainty relation is a set of $d^N$ inequalities of the following form
\begin{align}
P^{cert}(\rho,x)=\sum_{i=1}^N p(X_i)p(x_i|X_i)_{\rho}\leq C_{x}(\mathcal{O},\mathcal{P}),
\label{gFGU}
\end{align}
where $C_{x}(\mathcal{O},\mathcal{P})$ depends on the particular combination of measurement outcomes from a set of observables $\mathcal{O}=\{X_i\}$ and chosen with distribution function $\mathcal{P}=\{p(X_i)\}$. For the set of observables $\mathcal{O}=\{X_i\}$, the state which saturates Eq.(\ref{gFGU}), is a \textit{maximally certain state} for these observables. The quantity $C_{x}(\mathcal{O},\mathcal{P})$ captures the amount of uncertainty allowed in a particular physical theory. If $C_x(\mathcal{O},\mathcal{P})<1$ for any $x$, one cannot obtain any outcome with certainty. Later, in Ref.\cite{Rastegin2015} FUR were generalized for mutually unbiased bases (MUBs) in $d$ dimensional systems. For a set of $N$ mutual unbiased bases(MUBs) chosen with equal probability, the inequalities takes the following form \cite{Rastegin2015} 
\begin{align}\label{mub}
\frac{1}{N}\sum_{i=1}^N p(x_i|X_i)_{\rho}\leq \frac{1}{d}\left(1+\frac{d-1}{\sqrt{N}}\right).
\end{align}

Now, we will present FUR for a set of N arbitrary $d$-level observables.

\textbf{\textit{Result 1}}\label{result1} : 
\textit{For a set of N arbitrary observables in $dimension$ d, the FUR has the following form.} 
\begin{align}\label{fgu}
\frac{1}{N}\sum_{i=1}^N p(x_i|X_i)\leq \frac{1}{d}\left(1+\frac{(d-1)\sqrt{N +2\sum_{j>k=1}^N\cos(\theta_{jk})}}{N}\right),
\end{align}
\textit{where $\cos(\theta_{jk})$ is the angle between the bloch vectors corresponding to eigenvectors $\ket{x_j}$ and $\ket{x_k}$.}

\begin{proof}
To prove this, we need to find the state $\rho_{\max}$ which maximize the left hand side of Eq.(\ref{fgu}).
The eigenvectors $\ket{x_i}$ corresponding to eigenvalues $x_i$ and the state $\rho_{\max}$ can be expressed using Bloch vector representation as  \cite{Bertlmann_2008} 
\begin{align*}
\rho_{x_i}=\frac{1}{d}I +\vec{x_i}\cdot\vec{\Gamma}\hspace{0.1cm}\mbox{and}\hspace{0.1cm}
\rho_{\max}=\frac{1}{d}I +\vec{b}\cdot\vec{\Gamma},
\end{align*}
where $\vec{x_i}$ and $\vec{b}$ are the respective Bloch vectors and $\{\Gamma_i;i\in (0,...,d-1)\}$ are the generalised Gell-mann matrices in dimension $d$. The length of the Bloch vector in $d$ dimension  should be less than $\sqrt{(d-1)/2d}$, where the maximum length indicates pure states. The generalized Gell-Mann matrices are traceless, i.e. $\Tr (\Gamma_i) =0$ and orthogonal, i.e. $\Tr(\Gamma_i\Gamma_j)=2\delta_{ij}$ \cite{Bertlmann_2008}. 

Now, using the Bloch vector representation, we find that 
\begin{align*}
\frac{1}{N}\sum_{i=1}^N p(x_i|X_i)&=\frac{1}{N}\sum_{i=1}^{N}\Tr[\ketbra{x_i}{x_i}\rho_{\max}] \\
&= \frac{1}{N}\sum_{i=1}^{N}\Tr\left[\left(\frac{1}{d}I +\vec{x_i}.\vec{\Gamma}\right)\left(\frac{1}{d}I +\vec{b}\cdot\vec{\Gamma}\right)\right]\\
&= \frac{1}{N}\sum_{i=1}^{N}\left(\frac{1}{d}+2\vec{x_i}\cdot\vec{b}\right)\\
&= \frac{1}{d}+\frac{2}{N}\left(\sum_{i=1}^{N}\vec{x_i}\right)\cdot\vec{b}.
\end{align*}
It is straightforward to see that the quantity $\left(\sum_{i=1}^{N}\vec{x_i}\right)\cdot\vec{b}$ is maximum when $\vec{b}$ is collinear with  $ \sum_{i=1}^{N}\vec{x_i}$, i.e., $\vec{b}= \eta\sum_{i=1}^{N}\vec{x_i}$, where $\eta$ is the scaling factor.
For maximization, we have to find the appropriate value of $\eta$ such that $|\vec{b}|=\sqrt{\frac{d-1}{2d}}$, which implies that $\rho_{\max}$ must be a pure state. Since, $|\sum_{i=1}^{N}\vec{x_i}| = \sqrt{N'}\sqrt{\frac{d-1}{2d}}$, which yields $\eta=\frac{1}{\sqrt{N'}}$, where $N'=N +2\sum_{j>k=1}^N\cos(\theta_{jk})$. Thus, by substituting $\eta$, we find the bloch vector, $\vec{b}=\frac{1}{\sqrt{N'}}\sum_{i=1}^{N}\vec{x_i}$ and which gives upper bound for the considered FUR.
\end{proof}

These inequalities are tight for $d=2$, but not always tight for $d\geq 3$. This is so because not all the points on the surface of the $n^2-1$-dimensional hypersphere correspond to a valid pure state. As a corollary of our derivation fine-grained upper bound for MUBs can be reproduced using the following lemma.

\textbf{\textit{Lemma 1}} \label{lemma1}: \textit{The Bloch vectors belonging to $d$ dimensional mutually unbiased bases are orthogonal to each other.}
\begin{proof}
We notice that the overlap between two mutually unbiased state vectors is
\begin{align*}
\frac{1}{d} =\Tr[\left(\frac{1}{d}I +\vec{x_i}\cdot\vec{\Gamma}\right)\left(\frac{1}{d}I +\vec{x_j}\cdot\vec{\Gamma}\right)]=\frac{1}{d}+2\vec{x_i}\cdot\vec{x_j},
\end{align*} 
where we have used the tracelessness and orthogonality of the generalised Gell-mann matrices.
Therefore, we get $\vec{x_i}\cdot\vec{x_j}=0$. 
\end{proof}

Using the Lemma 1 in Eq.(\ref{fgu}), for any pair of mutually unbiased bases, $\cos(\theta_{jk})=0$ which  gives the Eq.(\ref{mub}). 
An example of the above inequality in qubit case, for measurements $\sigma_x$ and $\sigma_z$, is given by \cite{Oppenheim}
\begin{align*}
\frac{1}{2}p(x_{\sigma_x}|\sigma_x)+\frac{1}{2}p(x_{\sigma_z}|\sigma_z)\leq \frac{1}{2}\left(1+\frac{1}{\sqrt{2}}\right).
\end{align*}
\noindent The above inequality is saturated for all 4 possible vectors $\vec{x}\in \{x_{\sigma_x},x_{\sigma_z}\}$ and the maximally certain states are given by the eigenstates of $\frac{\sigma_x \pm \sigma_z}{\sqrt{2}}$. 
\subsection{Tight fine-grained uncertainty relations for two arbitrary observables in arbitrary dimension}

For two $d-$ dimensional observables $X_1$ and $X_2$, we can prove the following fine-grained inequalities corresponding to combination of outcomes  $x_1$ and $x_2$ respectively. 

\textbf{\textit{Result 2}} : The fine-grained inequality corresponding to obtaining outcomes  ${x_1}$ and ${x_2}$ by measurement of observables $X_1$ and $X_2$ respectively on the state $\rho$ has the following form
 
\begin{align}\label{tightfur}
\frac{1}{2}\Tr(\ketbra{x_1}{x_1}\rho)+\frac{1}{2}\Tr(\ketbra{x_2}{x_2}\rho)\leq \frac{1+|\braket{x_1}{x_2}|}{2}.
\end{align}

\begin{proof}
	Again we need to find the state $\rho$ which maximizes the left hand side term in Eq.(\ref{tightfur}).  For this, we will use the Landau-Pollak uncertainty which states that for two projectors $\ket{x_1}$ and $\ket{x_2}$ corresponding to outcomes $x_1$ and $x_2$ respectively, the following inequality exists
	\begin{align}\label{lp}
	\text{Arccos}\langle x_1 \rangle_{\rho}+\text{Arccos}\langle x_2 \rangle_{\rho} \geq  \text{Arccos}|\braket{x_1}{x_2}|.
	\end{align}
	
	Note that $\Tr(\ketbra{x_1}{x_1}\rho)=\langle x_1 \rangle_{\rho}^2 $ and similarly for $\Tr(\ketbra{x_2}{x_2}\rho)=\langle x_2 \rangle_{\rho}^2 $. From Eq.(\ref{lp}), we have $\text{Arccos}\langle x_2 \rangle_{\rho} \leq  \text{Arccos}|\braket{x_1}{x_2}|-\text{Arccos}\langle x_1 \rangle_{\rho}$.  We denote $\text{Arccos}\langle x_1 \rangle_{\rho} = \alpha$ and $\text{Arccos}|\braket{x_1}{x_2}|=\theta$ and substitute this inequality in the left-hand side term of Eq.(\ref{tightfur}).
	\begin{align*}
\left(\frac{1}{2}\Tr(\ketbra{x_1}{x_1}\rho)+\Tr(\ketbra{x_2}{x_2}\rho)\right)\leq \\ \frac{1}{2}\left( \cos^2 \alpha+\cos^2 (\theta-\alpha)\right).
	\end{align*}
	Now, finding the maximum of this expression is a simple optimization problem, which attains the maximum for $\alpha=\frac{\theta}{2}$ and give the upper bound in Eq.(\ref{tightfur}). Thus our inequality is proved.
\end{proof}
For MUBs the inequality in Eq.(\ref{tightfur}) becomes
\begin{align}\label{mubtight}
	\frac{1}{2}\Tr(\ketbra{x_1}{x_1}\rho)+\frac{1}{2}\Tr(\ketbra{x_2}{x_2}\rho)\leq \frac{1}{2}\left({1+\frac{1}{\sqrt{d}}}\right).
\end{align}

In the next section, we will present the states, which saturate the inequalities for MUBs when we connect FUR with random access code game.
\section{Violating  non-contextuality inequality with Fine-grained uncertainty }
In this section, we show how FUR determines the preparation contextuality of quantum theory. 
As previously stated,  there exist $d^N$ such inequalities for $N$ mutually unbiased observables Eq.(\ref{mubtight}). If we take the average over all such inequalities for $N=2$, we obtain 
\begin{align}\label{somub}
\frac{1}{2d^2}\sum_{x\in \{0,1,...,d-1\}^2}\sum_{i=1}^2 p(x_i|X_i)_{\rho_x}\leq \frac{1}{2}\left(1+\frac{1}{\sqrt{d}}\right),
\end{align}
where $x_i\in\{0,1,...,d-1\}$ are the measurement outcomes corresponding to observable $X_i$. If Alice encodes the classical string $x$ by preparing $\rho_x$, and sends to Bob, who measures $X_i$ to guess the $i^{th}$ bit of $x$, then L.H.S of inequality Eq.(\ref{somub}) becomes the success probability of $2\rightarrow1$ $d$ RAC game. Now, R.H.S of inequality Eq.(\ref{somub}) gives the quantum upper bound for the success probability of the game. Later we also show that such encoding and decoding scheme also respects the parity obliviousness condition, with respect to Bob's choice of measurements. Now, we state our result in terms of a theorem when Bob performs measurement with MUBs.

\subsection*{Preparation contextuality via 2$\rightarrow$ 1 $d$ PORAC game}
Using a 2$\rightarrow$ 1 $d$ RAC game one can demonstrate preparation contextuality on the basis of the following theorem.\\
\textbf{\textit{Theorem 3}}\label{theorem2}: \textit{In an RAC game, if Alice encodes the $2$-dit classical string $x$ in quantum states, which are maximally certain states for Bob's set of measurements (which are MUBs), then the preparation contextuality of quantum theory can be revealed. Moreover, this encoding-decoding strategy, guided by fine-grained uncertainty relations for MUBs satisfies Parity obliviousness condition, given by Eq.(\ref{oblc}) for  $2\rightarrow1$ $d$-PORAC.}

Demonstrating preparation contextual nature with an encoding and decoding scheme requires two things: 1) The RAC game should satisfy the parity obliviousness condition and 2) The success probability should be greater than that obtained in a non-contextual theory. First, we will show that our encoding and decoding scheme respects the parity obliviousness condition Eq.(\ref{oblc}), given a set of measurements performed by Bob.

\begin{proof}  $2\rightarrow 1$ $d$-RAC game has  $d$ sets of different parity and the number of classical message, $x=x_0 x_1$, in each set are $d$. We follow the encoding and decoding scheme presented in \cite{PhysRevLett.114.170502} for a $d-$level Quantum random access code game(QRAC), and show that their scheme is parity oblivious and can be derived from fine-grained inequalities for MUBs.  To detect the message $x_0x_1$ we use mutually unbiased basis given by the computational basis $\{\ket{p}\}_p$ and fourier basis $e_p=\frac{1}{\sqrt{d}}\sum_q^{d-1}\omega^{pq}\ket{q}$, where $\omega=\exp(\frac{2\pi i}{d})$. Alice encodes the classical signal $x_0x_1=00$ in the state  $\ket{\psi_{00}}=\frac{1}{N_d}(\ket{0}+\ket{e_0})$ where $N_d=\sqrt{2+\frac{2}{\sqrt{d}}}$. For the two projectors $\ket{0}$ and $\ket{e_0}$, this state  is the maximally certain state, i.e., it saturates the fine-grained inequality in Eq.(\ref{mubtight}). 
	Similarly, for other signals we use the encoding state as $\ket{\psi_{x_0x_1}}=X^{x_0}Z^{x_1}\ket{\psi_{00}}$, where $X=\sum_{q=0}^{d-1}\ketbra{q+1}{q}$ and $Z=\sum_{q=0}^{d-1}\omega^q\ketbra{q}{q}$ are the unitary operators.
	To learn about the first bit $x_0$ Bob will do the measurement in the computational basis and he will do measurement in the Fourier basis to learn about $x_1$. Given this encoding and decoding scheme, the success probability of Bob for determining the $x_0$ and $x_1$ bit is given by  
	\begin{align*}
	P_{x_0}(p)=|\braket{p}{\psi_{x_0x_1}}|^2=\frac{1}{N_d^2}\left|\delta_{x_0,p}+\frac{\omega^{x_1(p-x_0)}}{\sqrt{d}}\right|^2.
	\end{align*}
	
	\begin{align*}
	P_{x_1}(p)=|\braket{e_p}{\psi_{x_0x_1}}|^2=\frac{1}{N_d^2}\left|\omega^{-x_0x_1}\delta_{x_1,p}+\frac{\omega^{-px_0}}{\sqrt{d}}\right|^2.
	\end{align*}
	
	Bob's prediction is correct when either $p=x_0$ for $P_{x_0}$ and $p=x_1$ for $P_{x_1}$. In both the cases the success probability turns out to be $\frac{1}{2}\left(1+\frac{1}{\sqrt{d}}\right)$. Thus, it also saturates the fine grained inequality for the measurements in  $\{\ket{p}\}$ basis and $\{\ket{e_p}\}$ basis. 
	We note that since the success probability of $P_{x_0}$ is independent of the dit at position $x_1$ and similarly the success probability $P_{x_1}$ is independent of dit $x_0$, our encoding and decoding scheme is parity oblivious in this scenario.
\end{proof}

Next, we will show that the success probability in our encoding and decoding scheme exceeds the non-contextual bound of a PORAC.
\begin{proof}	
The maximum success probability of the $2\rightarrow1$ $d$-level RAC game in quantum theory is exactly the  R.H.S of the Eq.(\ref{somub}). On comparing the upper bound of $N\rightarrow1$ $d$ PORAC game  with that of FUR, we find that $\frac{1}{d}\left(1+\frac{d-1}{\sqrt{2}}\right)\geq\frac{2+d-1}{2d}=\frac{1}{d}\left(1+\frac{d-1}{2}\right)$. Therefore, we have obtained a violation of the preparation non-contextuality inequality. 
\end{proof} 

It should be noted that a full set of MUBs for an arbitrary dimension $d$ is not known. But in the 2$\rightarrow$ 1 $d$ PORAC game, we need only 2 such observables of dimension $d$ for our scheme to work.

\textit{Example-1} First we present the simplest example of a $2\rightarrow1$, 2-PORAC. Although this has been presented earlier \cite{spekkenpom}, we only highlight how the fine-grained uncertainty relations comes in the picture. The classical signal \{00,01,10,11\} are encoded in the states with Bloch vectors $\left(0,\pm\frac{1}{\sqrt{2}},\pm\frac{1}{\sqrt{2}}\right)$, because for $\sigma_x$ and $\sigma_y$ these states saturate the fine-grained uncertainty relation. To decode the signal Bob measures with $\sigma_x$ to measure the first bit and with $\sigma_y$ to measure the second bit. Using this method he detects the correct signal with probability $\frac{1}{2}\left(1+\frac{1}{\sqrt{2}}\right)=0.8535553\geq \frac{n+1}{2n}=\frac{3}{4}$, and thus violates the inequality in Eq.(\ref{ncie}). The parity obliviousness condition is also respected, since the parity 0 and 1 states are represented by the same density matrix operator,i.e., $\frac{1}{2}\rho_{00}+\frac{1}{2}\rho_{11}=\frac{1}{2}\rho_{10}+\frac{1}{2}\rho_{01}=\frac{I}{2}$. Thus, by using the fine grained uncertainty relation we obtain a violation of preparation non-contextuality.

\textit{Example-2} In \cite{PhysRevLett.114.170502}, the authors have found a violation of $2\rightarrow1$ 3-PORAC game analytically as well as numerically. Although, they didn't mention it as a parity oblivious game. The analytical value of success probability is $\frac{1}{2}\left(1+\frac{1}{\sqrt{3}}\right)=0.788675$. As we have shown, their encoding and decoding scheme is parity oblivious also.
 
\subsection*{Parity obliviousness in $3\rightarrow 1$ 2 PORAC game}
In this section, we use the fine-grained inequality in Eq.(\ref{mub}) for demonstrating preparation contextuality. Since this inequality is tight only for $d=2$ we limit ourselves to that. In the  $3\rightarrow 1,  2-$PORAC game, Alice encodes the classical signal \{000,001,010,011,100,101,110,111\} in qubit quantum states and sends them to Bob. Following the fine-grained inequality in Eq.(\ref{fgu}), if Alice encodes them states with Bloch vectors $ \left(\pm\frac{1}{\sqrt{3}},\pm\frac{1}{\sqrt{3}},\pm\frac{1}{\sqrt{3}}\right)$, which saturate the fine-grained uncertainty for 3 observables $\sigma_x$, $\sigma_y$ and $\sigma_z$ with mutually unbiased bases. Bob employs $\sigma_x$,$\sigma_y$ and $\sigma_z$ operators to detect the first, second and third bit respectively and obtains correct signal with probability $\frac{1}{2}\left(1+\frac{1}{\sqrt{3}}\right)=0.788675\geq \frac{n+1}{2n}=\frac{2}{3}$. It has been shown that this encoding scheme is also parity oblivious \cite{spekkenpom}. Moreover, this is the optimal success probability for this game Refs.\cite{Chailloux_2016,PhysRevA.98.032110}.

Now, we find an upper bound of the quantum violation of preparation non-contextuality inequality.

\textbf{\textit{Theorem 4}}\label{theorem4}: \textit{Encoding-decoding strategy based on FUR for MUBs gives the upper bound of quantum violation of preparation the non-contextuality inequality, specifically, it gives an upper bound of success probability of quantum theory in the $N\rightarrow1$ $d$-PORAC game, in which decoding is done using rank-one projective measurements.}\\
We prove Theorem 4 in Appendix \ref{appendix2}.

\section{Conclusion}
The optimal success probability of certain communication games reveal the fundamental limitations of different operational theories. Quantum advantage of random access code game with the additional constraint of parity obliviousness asserts that quantum theory is preparation contextual. Here, we show that the success probability of a parity oblivious RAC game is determined by the amount of fine-grained uncertainty for Bob's choice of rank-one projective measurements. To show this, we have derived an upper bound for fine-grained uncertainty relations of  $N$ arbitrary observables of dimension $d$. In addition, we have also found tight fine-grained inequalities for two observable, which provide optimal encoding and decoding strategy for $2\rightarrow1$ d-PORAC. Subsequently, we find analytically, the quantum violation of the preparation contextuality inequality for the $ 2\rightarrow1 $ $d$-PORAC game. Some partial results of optimal violations were known up to a few dimension with the help of numerical methods i.e., semidefinite programming \cite{Ambainis2019}. Our results are derived under the condition that the dimension of the resource states corresponding to $d$-PORAC game is also $d$ in classical or quantum theory.  In future, one can try to find the violation of preparation non-contextuality for $N\rightarrow1$ d-PORAC games, for $N>2$ also.

\section{Acknowledgement}
GS would like to acknowledge the Department of Atomic Energy, Govt. of India, for providing research fellowships. The research of GS was also supported in part by the INFOSYS scholarship for senior students.

\section*{Supplementary Material}
\appendix
\section{}
\textbf{\textit{Lemma 2}}\label{lemma2}: \textit{The sum of the Bloch vectors corresponding to eigenvectors of an observable is zero.}

\begin{proof}
 The eigenvectors $\ket{v_i}$ of an observable $\mathcal{O}$ satisfy $\sum_i\ketbra{v_i}{v_i}=I$. In terms of bloch vectors $\vec{b}$, one can write
 \begin{align*}
  \ketbra{v_i}{v_i}=\frac{1}{d}I +\vec{b_i}\cdot\vec{\Gamma}.
 \end{align*}
By taking a sum over all the eigenvectors, we get $\sum_i\ketbra{v_i}{v_i}= I+\left(\sum_i\vec{b_i}\right)\cdot\vec{\Gamma}=I$, which gives $\sum_i\vec{b_i}=0$.
\end{proof}

\section{}\label{appendix2}
Here, we will prove Theorem 4, and find the maximal success probability of a $N \rightarrow 1$ d-PORAC game, over all possible measurement settings. The success probability of the N $\rightarrow$1 d-PORAC is given by 
\begin{align*}
p_{succ}=\frac{1}{d^NN}\sum_{x\in \{0,1,...,d-1\}^N}\sum_{i=1}^N p(x_i|X_i)_{\rho},
\end{align*}
where we have not specified the measurement setting chosen by Bob. For any arbitrary measurement performed by Bob, $p(x_i|X_i)_{\rho}=\frac{1}{d}+2\vec{x_i}\cdot \vec{b}$, where $\vec{x_i}$ is the bloch vector corresponding to the outcome $x_i$, where $\vec{b}$ is the bloch vector of encoding state. For optimal encoding it will depend on the $\vec{x_i}$'s, which we will prove now. Substituting this probability in the above equation we get
\begin{align}\label{psucc}
 p_{succ}&=\frac{1}{d^NN}\sum_{x\in \{0,1,...,d-1\}^N}\sum_{i=1}^N \left(\frac{1}{d}+2\vec{x_i}\cdot \vec{b}\right)\nonumber \\ 
 &=\frac{1}{d}+\frac{2}{Nd^N}\sum_{x\in \{0,1,...,d-1\}^N}\sum_{i=1}^N\vec{x_i}\cdot \vec{b} \nonumber \\ 
 &=\frac{1}{d}+\frac{2 \:\Phi(\vec{X},\vec{b})}{Nd^N},
\end{align}
where $\Phi(\vec{X},\vec{b})=\sum_{x\in \{0,1,...,d-1\}^N}\sum_{i=1}^N\vec{x_i}\cdot \vec{b}$. To get the optimal success probability, we need to maximize $\Phi(\vec{X},\vec{b})$ over all possible measurements $X_i$ and encodings $\vec{b}$. We denote the maximum value as $\Phi(N)$.
\begin{align*}
 \Phi(N)=\max_{\vec{X},\vec{b}}\Phi(\vec{X},\vec{b})=\max_{\vec{X}}\sum_{x\in \{0,1,...,d-1\}^N}\max_{\vec{b}}\vec{b}\cdot \sum_i\vec{x_i}
\end{align*}
The second maximization can be easily done by chosing $\vec{b}$ in the direction of $\sum_i\vec{x_i}$, so that $\vec{b}\cdot \sum_i\vec{x_i}=\sqrt{\frac{d-1}{2d}}||\sum_i\vec{x_i}||$. Then, 
\begin{align*}
 \Phi(N)&=\max_{\vec{X},\vec{b}}\Phi(\vec{X},\vec{b})=\sqrt{\frac{d-1}{2d}}\max_{\vec{X}}\sum_{x\in \{0,1,...,d-1\}^N}||\sum_i\vec{x_i}||.
\end{align*}

To find the value of $\Phi(N)$, we use the following Lemma

\textbf{\textit{Lemma 3}}: \textit{For vectors $\vec{x_i}$, we have, $\sum_{x\in \{0,1,...,d-1\}^N}||\sum_{i=1}^N \vec{x_i}||^2=\frac{(d-1)}{2d}N d^N$.}

\begin{proof}
 We prove this Lemma by induction. For $N=1$, we have 
 \begin{align*}
  \sum_{x\in \{0,1,...,d-1\}}||\vec{x_1}||^2=\frac{d(d-1)}{2d}.
 \end{align*}

 Assuming that our lemma holds for $N=m$, then for $N=m+1$ we have 
 \begin{align*}
  &\sum_{x\in \{0,1,...,d-1\}^{m+1}}||\sum_{i=1}^{m+1} \vec{x_i}||^2 \\&=\sum_{x\in \{0,1,...,d-1\}^{m+1}}||\vec{x_1}+ \vec{x_2}+...+\vec{x_{m+1}}||^2.
 \end{align*}
By summing over the $m+1$ index, we get
 \begin{align*}
  \sum_{x\in \{0,1,...,d-1\}^m}\big(||\vec{x_1}+...+\vec{x_{m}}||^2+||\vec{x_{m+1}}||^2\\+2\vec{x_{m+1}}(\vec{x_1}+...+\vec{x_{m}})\big).
\end{align*}

%
By using Lemma 2, we note that 
\begin{align*}
\sum_{x\in \{0,1,...,d-1\}^N}2\vec{x_{m+1}}(\vec{x_1}+\vec{x_2}+...+\vec{x_{m}})=0.
\end{align*}

Since we have assumed that the lemma holds for $N=m$, the above expression simplifies to $d(m\cdot d^m+d^m)\frac{(d-1)}{2d}$=$\frac{(d-1)}{2d}(m+1)d^{m+1}$.
\end{proof}

Now, $\Phi(N)$ can be seen as an inner product between $\sum_{x\in\{0, 1,...,d-1\}^N}\sum_{i=1}^N \vec{x_i}$ and the vector $(1,1,...,1)\in \mathcal{R}^{d^N}$, hence we can apply the Cauchy -Schwarz inequality to get an upper bound on $\Phi(N)$, so that
\begin{align*}
 \Phi(N)\leq \sqrt{\frac{(d-1)}{2d}}\sqrt{d^N}\sqrt{\frac{(d-1)}{2d}N d^N}=\frac{\sqrt{N}(d-1)}{2d} d^N.
\end{align*}

By substituting $\Phi(N)$ in Eq.(\ref{psucc}), we get the maximum success probability as
\begin{align*}
 p_{succ}=\frac{1}{d}\left(1+\frac{d-1}{\sqrt{N}}\right).
\end{align*}

\textit{ Note:}  We are finding an upper bound of a $N\rightarrow1$  d-RAC game, based on projective measurements. Therefore, to find quantum upper bound we consider FUR inequalies invloving projective measurements. The maximum success probability for $2\rightarrow1$ RAC game is obtained by encoding the signal in a 2 dimensional quantum state. Therefore, for a $N\rightarrow1$  d level RAC game also, we have restricted the dimension of the encoding state(classical/quantum) to be equal to dimension $d$.
\end{document}